\documentclass[12pt,twoside]{article}

\evensidemargin=\oddsidemargin
\def\Title#1#2#3{%
    \baselineskip=18pt
    \begin{center}
          {\large\bf{#1} \\ }
          \bigskip\bigskip
          {#2} \\
          {#3} \\
    \end{center}}
\long\def\Abstract#1{%
         \bigskip
         \parbox{0.93\textwidth}{%
                 \begin{center}
                       {\bf Abstract} \\
                 \end{center}
                 \medskip{\baselineskip=14pt #1}
                 \vss}
         \bigskip}

\begin{document}

\Title{Hamiltonian dynamics in extended phase space for gravity\\
and its consistency with Lagrangian formalism:\\
a generalized spherically symmetric model as an example}%
{T. P. Shestakova}%
{Department of Theoretical and Computational Physics,
Southern Federal University,\\
Sorge St. 5, Rostov-on-Don 344090, Russia \\
E-mail: {\tt shestakova@sfedu.ru}}

\Abstract{Among theoretical issues in General Relativity the problem of constructing its Hamiltonian formulation is still of interest. The most of attempts to quantize Gravity are based upon Dirac generalization of Hamiltonian dynamics for system with constraints. At the same time there exists another way to formulate Hamiltonian dynamics for constrained systems guided by the idea of extended phase space. We have already considered some features of this approach in the previous MG12 Meeting by the example of a simple isotropic model. Now we apply the approach to a generalized spherically symmetric model which imitates the structure of General Relativity much better. In particular, making use of a global BRST symmetry and the Noether theorem, we construct the BRST charge that generates correct gauge transformations for all gravitational degrees of freedom.} 

\bigskip

As a rule, Hamiltonian formulation for gravity is constructed following to the Dirac scheme \cite{Dirac1,Dirac2} and is a starting point for most attempts to quantize gravity. However, there are some reasons to doubt that Dirac Hamiltonian formulation for gravitational theory \cite{Dirac3} can be thought as an equivalent one to the original General Relativity. Indeed, in Einstein formulation of General Relativity $g_{0\mu}$ components of metric tensor ($\mu=0, 1, 2, 3$) are treated on an equal basis with the rest of components, $g_{ij}$ ($i, j=1, 2, 3$). The theory is invariant under gauge transformations that touch all metric components. In the Dirac approach only $g_{ij}$ components with their conjugate momenta are included into phase space, the transformations for this variables are generated by constraints. In fact, the original General Relativity and the Dirac formulation are theories with different groups of transformations. We can think of it as a considerable mathematical indication that Lagrangian and Hamiltonian formalism appear to be non-equivalent for the full theory of gravity. Moreover, the algebra of constraints in Dirac phase space does depend on parametrization of gravitational variables. It creates a serious obstacle to find an algorithm to construct a generator that would give correct transformations for all gravitational variables in the limits of the Dirac approach \cite{Shest1}. The situation is essentially the same in the Batalin -- Fradkin -- Vilkovisky (BFV) approach \cite{BFV1,BFV2,BFV3}. The central part in the BFV approach is given to the BRST charge. In the BFV approach the form of the BRST charge is determined by constraints algebra. It is not surprising that it does not generates correct gauge transformations for all gravitational degrees of freedom. At the same time, the existing of a global BRST symmetry enables us to propose another method based upon the Noether theorem and the equivalence of Lagrangian dynamics and Hamiltonian dynamics in extended phase space.

Our purpose is to construct Hamiltonian dynamics in extended phase space for a generalized spherically symmetric gravitational model which is free from the shortcomings mentioned above and can be proved to be completely equivalent to its Lagrangian formulation. We shall follow to the ADM parametrization \cite{ADM}. Under the condition of spherical symmetry the metric is reduced to
\begin{eqnarray}
\label{mod.int}
ds^2&=&\left[-N^2(t,r)+(N^r(t,r))^2V^2(t,r)\right]dt^2+2N^r(t,r)V^2(t,r)dt dr\nonumber\\
   &+&V^2(t,r)dr^2+W^2(t,r)\left(d\theta^2+\sin^2\theta d\varphi^2\right).
\end{eqnarray}
where $N^r=N^1$ is the only component of the shift vector. In this model we have two gauge variables $N$ and $N^r$ which are fixed by two gauge conditions in differential form which introduce missing velocities into the effective Lagrangian:
\begin{equation}
\label{diff.conf}
\dot N=\frac{\partial f}{\partial V}\dot V+\frac{\partial f}{\partial W}\dot W;\qquad
\dot N^r=\frac{\partial f^r}{\partial V}\dot V+\frac{\partial f^r}{\partial W}\dot W.
\end{equation}

We shall consider the Faddeev -- Popov effective action including gauge and ghost sectors as it appears in the path integral approach to gauge field theories. To get field equations it is much easier to make use of the Lagrangian which is quadratic in first derivatives of metric components and ghosts and can be obtained by omitting total derivatives.
\begin{eqnarray}
\label{full-act2}
S&=&\int dt\int\limits_0^{\infty}dr\left(\frac{\dot V\dot W W}N
  +\frac{V\dot W^2}{2N}-\frac{N'W'W}V-\frac{N(W')^2}{2V}
  -\frac{NV}2-\frac{W'\dot WVN^r}N\right.\nonumber\\
 &-&\frac{W W'\dot V N^r}N-\frac{W\dot W V'N^r}N
  -\frac{W\dot W V(N^r)'}N+\frac{W W'V'(N^r)^2}N
  +\frac{W W'V N^r(N^r)'}N\nonumber\\
 &+&\frac{(W')^2 V(N^r)^2}{2N}
  +\pi_N\left(\dot N-\frac{\partial f}{\partial V}\dot V-\frac{\partial f}{\partial W}\dot W\right)
  +\pi_{N^r}\left(\dot N^r-\frac{\partial f^r}{\partial V}\dot V
   -\frac{\partial f^r}{\partial W}\dot W\right)\nonumber\\
 &+&\dot{\bar\theta}_0\theta^r\left(N'-\frac{\partial f}{\partial V}V'
   -\frac{\partial f}{\partial W}W'\right)
  +\dot{\bar\theta}_0\left(N\dot\theta^0-NN^r(\theta^0)'
   -\frac{\partial f}{\partial V}VN^r(\theta^0)'\right.\nonumber\\
 &-&\left.\frac{\partial f}{\partial V}V(\theta^r)'\right)
   +\dot{\bar\theta}_r\left[N^r\dot\theta^0-\left(\frac{N^2}{V^2}+(N^r)^2\right)(\theta^0)'+\dot\theta^r
   -N^r(\theta^r)'+(N^r)'\theta^r\right.\nonumber\\
 &-&\left.\left.\frac{\partial f^r}{\partial V}\left(VN^r(\theta^0)'+V(\theta^r)'+V'\theta^r\right)
   -\frac{\partial f^r}{\partial W}W'\theta^r\right]\right)
\end{eqnarray}

Variation of the effective action yields the Einstein equations with additional terms resulting from the gauge-fixing and ghost parts of the action, as well as ghost equation and gauge conditions. Introducing of the missing velocities by means of the differential form of gauge conditions enables us to construct a Hamiltonian by the usual rule
\begin{equation}
\label{Ham1}
H=\int\limits_0^{\infty}dr\left(P_N\dot N+P_{N^r}\dot N^r+P_V\dot V+P_W\dot W
   +\bar P_{\theta^0}\dot\theta^0+\dot{\bar\theta}_0P_{\bar\theta_0}
   +\bar P_{\theta^r}\dot\theta^r+\dot{\bar\theta}_rP_{\bar\theta_r}-L\right)
\end{equation}
and derive Hamiltonian equations in extended phase space. An explicit form of the Hamiltonian function, Lagrangian and Hamiltonian equations can be found in \cite{Shest2}.

The first step to construct the BRST charge according to the Noether theorem is to find a BRST invariant form of the action. In the case of gravity we deal with space-time symmetry, and we should take into account explicit dependence of the Lagrangian and the measure on space-time coordinates. We should use a gauge invariant form of gravitational action and add to the action the following term containing only full derivatives and not affecting Lagrangian equations:
\begin{eqnarray}
\label{add-act}
S_{(add)}&=&\int dt\int\limits_0^{\infty}dr
   \left(\frac d{dt}\left[\bar\theta_0\left(\dot N-\frac{\partial f}{\partial V}\dot V
     -\frac{\partial f}{\partial W}\dot W\right)\theta^0\right]\right.\nonumber\\
&+&\frac d{dr}\left[\bar\theta_0\left(\dot N-\frac{\partial f}{\partial V}\dot V
     -\frac{\partial f}{\partial W}\dot W\right)\theta^r\right]
     +\frac d{dt}\left[\bar\theta_r\left(\dot N^r-\frac{\partial f^r}{\partial V}\dot V
     -\frac{\partial f^r}{\partial W}\dot W\right)\theta^0\right]\nonumber\\
&+&\left.\frac d{dr}\left[\bar\theta_r\left(\dot N^r-\frac{\partial f^r}{\partial V}\dot V
     -\frac{\partial f^r}{\partial W}\dot W\right)\theta^r\right]\right).
\end{eqnarray}

The resulting BRST charge is
\begin{eqnarray}
\label{BRST}
\Omega&=&\int\!dr\left[-{\cal H}\theta^0-P_V V'\theta^r
   -P_N\frac{\partial f}{\partial V}V'\theta^r
   -P_{N^r}\frac{\partial f^r}{\partial V}V'\theta^r
   -P_W W'\theta^r\right.\nonumber\\
&-&P_N\frac{\partial f}{\partial W}W'\theta^r
   -P_{N^r}\frac{\partial f^r}{\partial W}W'\theta^r
   -P_V V N^r(\theta^0)'
   -P_N\frac{\partial f}{\partial V} V N^r(\theta^0)'\nonumber\\
&-&P_{N^r}\frac{\partial f^r}{\partial V} V N^r(\theta^0)'
   -P_V V(\theta^r)'-P_N\frac{\partial f}{\partial V} V(\theta^r)'
   -P_{N^r}\frac{\partial f^r}{\partial V} V(\theta^r)'\nonumber\\
&-&\left.\bar P_{\theta^0}(\theta^0)'\theta^r
   -\bar P_{\theta^r}(\theta^r)'\theta^r
   -P_N P_{\bar\theta_0}-P_{N^r}P_{\bar\theta_r}
   -\frac{N W W'(\theta^0)'}V\right],
\end{eqnarray}
$\cal H$ is a Hamiltonian density in (\ref{Ham1}). The group of transformations generated by (\ref{BRST}) includes the group of gauge transformations for all gravitational degrees of freedom. The proposed approach is applicable to any constrained system and can be considered as a preliminary step to subsequent quantization of the model.


\small

\begin{thebibliography}{99}
\itemsep=-5pt
\bibitem{Dirac1}
P. A. M. Dirac,
 {\it Can. J. Math.\/} {\bf 2} (1950), P. 129--148.
\bibitem{Dirac2}
P.~A.~M. Dirac,
 {\it Proc. Roy. Soc.\/} {\bf A246} (1958), 326--332.
\bibitem{Dirac3}
P. A. M. Dirac,
 {\it Proc. Roy. Soc.\/} {\bf A246} (1958), P. 333--343.
\bibitem{Shest1}
T. P. Shestakova,
 {\it Class. Quantum Grav.\/} {\bf 28} (2011), 055009.
\bibitem{BFV1}
E. S. Fradkin and G. A. Vilkovisky,
 {\it Phys. Lett.\/} {\bf B55} (1975), P. 224--226.
\bibitem{BFV2}
I. A. Batalin and G. A. Vilkovisky,
 {\it Phys. Lett.\/} {\bf B69} (1977), P. 309--312.
\bibitem{BFV3}
E. S. Fradkin and T. E. Fradkina,
 {\it Phys. Lett.\/} {\bf B72} (1978), P. 343--348.
\bibitem{ADM}
R. Arnowitt, S. Deser and C. W. Misner,
 ``The Dynamics of General Relativity'',
 in: {\it Gravitation, an Introduction to Current Research\/},
 ed. by L. Witten, John Wiley \& Sons, New York (1962), P. 227--265.
\bibitem{Shest2}
T. P. Shestakova,
  ``Generalized spherically symmetric gravitational model: Hamiltonian dynamics in extended phase space and BRST charge'', E-print arXiv: gr-qc1302.4875.
\end{thebibliography}
\end{document}